\documentclass[reprint,twocolumn,superscriptaddress,secnumarabic,amssymb, nobibnotes, aps,prb]{revtex4-2}
 
\usepackage{extarrows}
\usepackage{amsmath}    % need for subequations
\usepackage{graphicx}    % need for figures
\usepackage{verbatim}    % useful for program listings
\usepackage{color}           % use if color is used in text
\usepackage{subfigure}   % use for side-by-side figures
\usepackage{braket}
\usepackage{hyperref}    % use for hypertext links, including those to external documents 
\usepackage{verbatim}  % and URLs
\usepackage{here}
\usepackage{mathrsfs}
\usepackage[english]{babel}
\usepackage[utf8x]{inputenc}
\usepackage[T1]{fontenc}
\usepackage{comment}
\usepackage[version=4]{mhchem}

\begin{document}

\title{Atomic multiplet and charge-transfer screening effects in 1$s$ and 2$p$ core-level x-ray photoelectron spectra of early 3$d$ transition-metal oxides}
%\\ 
%\tc{blue}{Core-level photoelectron spectra of early transition-metal oxides: Atomic multiplet effect vs charge-transfer screening}}

\author{Tatsuya~Yamaguchi}
\thanks{T.Y. and K.H. contributed equally to this work.}
\affiliation{Department of Physics and Electronics, Graduate School of Engineering, Osaka Metropolitan University 1-1 Gakuen-cho, Nakaku, Sakai, Osaka 599-8531, Japan.}

\author{Keisuke~Higashi}
\thanks{T.Y. and K.H. contributed equally to this work.}
\affiliation{Department of Physics and Electronics, Graduate School of Engineering, Osaka Metropolitan University 1-1 Gakuen-cho, Nakaku, Sakai, Osaka 599-8531, Japan.}

\author{Anna~Regoutz}
\affiliation{Department of Chemistry, University College London, 20 Gordon Street, London WC1H 0AJ, United Kingdom.}%

%\author{ Louis~F.~J.~Piper}
%\affiliation{WMG, University of Warwick, Coventry CV4 7AL, United Kingdom.}
%\affiliation{The Faraday Institution, Quad One, Harwell Campus, Didcot OX11 0RA, United Kingdom.}%

\author{Yoshihiro Takahashi}
\affiliation{Department of Physics and Electronics, Graduate School of Engineering, Osaka Metropolitan University 1-1 Gakuen-cho, Nakaku, Sakai, Osaka 599-8531, Japan.}

\author{Masoud~Lazemi}
\affiliation{Materials Chemistry and Catalysis, Debye Institute for Nanomaterials Science, Utrecht University, Universiteitsweg 99, 3584 CG, Utrecht, The Netherlands.}
%\affiliation{Inorganic Chemistry \& Catalysis, Debye Institute for Nanomaterials Science, Utrecht University, Universiteitsweg 99, 3584 CG, Utrecht, The Netherlands.}

\author{Qijun~Che}
\affiliation{Materials Chemistry and Catalysis, Debye Institute for Nanomaterials Science, Utrecht University, Universiteitsweg 99, 3584 CG, Utrecht, The Netherlands.}
%\affiliation{Inorganic Chemistry \& Catalysis, Debye Institute for Nanomaterials Science, Utrecht University, Universiteitsweg 99, 3584 CG, Utrecht, The Netherlands.}

\author{Frank~M.~F.~de~Groot}
\affiliation{Materials Chemistry and Catalysis, Debye Institute for Nanomaterials Science, Utrecht University, Universiteitsweg 99, 3584 CG, Utrecht, The Netherlands.}
%\affiliation{Inorganic Chemistry \& Catalysis, Debye Institute for Nanomaterials Science, Utrecht University, Universiteitsweg 99, 3584 CG, Utrecht, The Netherlands.}

\author{Atsushi~Hariki}
\affiliation{Department of Physics and Electronics, Graduate School of Engineering, Osaka Metropolitan University 1-1 Gakuen-cho, Nakaku, Sakai, Osaka 599-8531, Japan.}
%\email{hariki@omu.ac.jp}

\date{\today}

\begin{abstract}
We present a comparative analysis of 1$s$ and 2$p$ core-level hard x-ray photoelectron spectroscopy (HAXPES) spectra of metallic VO$_2$ and CrO$_2$. Even though the V 1$s$ and 2$p$ spectra in VO$_2$ show similar line shapes except for the absence/presence of a spin-orbit  splitting in the core levels, the Cr 1$s$ and 2$p$ spectra exhibit distinct main-line shapes. The experimental HAXPES spectra are analyzed by the Anderson impurity model based on the density functional theory %\tc{blue}{(LDA)} 
+ dynamical mean-field theory 
%\tc{blue}{(DMFT)} 
and a conventional MO$_6$ cluster model. We examine the interplay between the formation of the intraatomic multiplet between a core hole and valence electrons and the charge-transfer effect on the chemical bonding followed by the 1$s$ and 2$p$ core electron excitations. We demonstrate the advantage of 1$s$ HAXPES to the routinely employed 2$p$ one for distinguishing contributions of a metal-ligand and metallic screening from a state near the Fermi level in metallic early 3$d$ transition-metal oxides.
\end{abstract}

\maketitle

%%%%%%%%%%%%%%%%%%%%%%%%%%%%%%%%%%%%%%%%%%%%%%%%%%%%%%%%%
\section{Introduction}
%%%%%%%%%%%%%%%%%%%%%%%%%%%%%%%%%%%%%%%%%%%%%%%%%%%%%%%%%

Transition metal oxides (TMOs) exhibit a variety of functional properties ranging from unconventional superconductivity with high $T_{\rm c}$ to exotic phase transitions, including metal-insulator transitions~\cite{imada98,khomskii14}.
%These intriguing properties arise from the complex interplay between the formation of intraatomic multiplets and inter-atomic hybridization. 
These properties often arise from the complex interplay between local electron-electron interaction and interatomic hybridization. The formation of intraatomic multiplet introduces additional complexity in multiorbital cases, resulting in nontrivial correlated states, even in metallic systems, such as Hund's metal~\cite{Medici11,Stadler15,Georges13}.
Core-level x-ray photoelectron spectroscopy (XPS) is a powerful tool to investigate TMOs~\cite{groot_kotani}. 
In XPS, a highly-localized core hole created by x-ray irradiation strongly interacts with the $d$ electrons on the same transition metal (TM) ion,
leading to %a formation of a bound exciton with 
atomic multiplet fine structures %which are reflected
in the XPS spectrum.
Besides, the sudden appearance of the core hole induces a dynamical charge response from surrounding ions to the excited ion, traditionally called charge-transfer (CT) screening~\cite{veenendaal06,veenendaal93,ghiasi2019}. The CT screening often generates satellite peaks in the core-level spectra, providing valued insights into the chemical bonding properties.
%\textcolor{black}{Furthermore, **** value of TM-TM nonlocal CT ****.}
The development of hard x-ray photoelectron spectroscopy (HAXPES) has enabled the exploration of bulk properties and deeper core levels of materials compared to conventional soft x-rays~\cite{taguchi13,Taguchi16book,BORGATTI201695,Kalha_2021_HAXPES_review,kobayashi09,groot_kotani}
%~\cite{taguchi13,Taguchi16book,BORGATTI201695,Kalha_2021_HAXPES_review,kobayashi09,groot_kotani}. 

Recently, TM 1$s$ HAXPES spectra were examined for insulating late (NiO, Fe$_2$O$_3$)~\cite{ghiasi2019,zuazo18} and early 3$d$ TMOs (TiO$_2$, SrTiO$_3$)~\cite{Woicik20,Woicik15,Hariki22}. 
For the band-insulating Ti oxides with a $d^0$ configuration in the formal valence, Ti 1$s$ HAXPES revealed low-energy satellites that are not seen in the Ti 2$p_{3/2}$ one, since the satellites are positioned at the same binding energies $E_B$ as the Ti 2$p_{1/2}$ spin-orbit-partner. % with larger spectral intensities~\cite{Hariki22}. 
In Fe$_2$O$_3$ (Mott insulator) and NiO (CT insulator), the 1$s$ spectra clearly exhibit CT screening features, including a nonlocal CT channel from the distant Ni or Fe ions beyond the nearest-neighboring ligands~\cite{veenendaal93,veenendaal06,Taguchi16book,hariki17}, despite a larger lifetime broadening than the 2$p$ core-level excitation~\cite{ghiasi2019,zuazo18}. 
These findings suggest the 1$s$ core-level excitation as a complementary tool to the routinely used 2$p$ excitation for studying 3$d$ TMOs. 
\textcolor{black}{However, the previous studies have exclusively focused on the insulators. In metallic TMOs, 2$p$ HAXPES spectra often exhibit a metallic nonlocal screening feature (edge singularity) from a conducting state at the Fermi level, providing insights into the correlated metallic properties~\cite{horiba04,taguchi10,taguchi05,eguchi08,Suga09,Paez2020,Taguchi16book,Taguchi05b,Haverkort14,Cornaglia07}. Yet, the metallic screening effect in 1$s$ spectra has not been explored.}

Another aspect in the 1$s$ core-level excitation is the absence of the core-valence Coulomb multiplet effect, \textcolor{black}{which would eliminate fine features and thus allow an easy access to the valence and chemical bonding information.}
%without encountering the complexities in the fitting analysis of the core-level spectra.
However, this potential benefit of 1$s$ HAXPES is not emphasized in the previous studies,  
\textcolor{black}{primarily because the TM 3$d$ shell is almost empty in SrTiO$_3$ and TiO$_2$, and largely occupied in NiO.} 
In such cases, the core-valence multiplet effect is weak in the 2$p$ XPS spectra.
\textcolor{black}{Thus, survey for other TM oxides with rich core-valence multiplet effect in the 2$p$ spectra will be valuable to examine the benefit of the multiplet-free 1$s$ excitation.} 
%, however, was not explored for unveiling intrinsic spectral features.
%Furthermore, the visibility of metallic nonlocal CT screening features in the 1$s$ spectrum needs to be examined in TMOs as the aforementioned examples in recent studies~\cite{ghiasi2019,Hariki22,zuazo18} are all insulators.

In this study, \textcolor{black}{we present a systematic comparison of 1$s$ and 2$p$ HAXPES spectra of metallic vanadium dioxide (VO$_2$) and chromium dioxide (CrO$_2$). We examine the sensitivity of the 1$s$ excitation to the metallic nonlocal screening and demonstrate difference in the 1$s$ and 2$p$ HAXPES spectra, where the core-valence atomic multiplet effect is not negligible in the 2$p$ excitation.}
VO$_2$ exhibits a metallic behavior in the high-temperature rutile structure phase.
The V 2$p$ HAXPES spectrum displays a shoulder feature associated with the V 2$p_{3/2}$ main line due to a metallic CT screening~\cite{eguchi08,Suga09,Paez2020}.
In the previous studies~\cite{eguchi08,Suga09}, the V 1$s$ HAXPES spectrum exhibits a similar shoulder feature.
%although a theoretical analysis of the 1$s$ data was not performed so far.
%in the high-temperature metallic phase despite the nonlocal-screening feature is suppressed in the low-temperature insulating phase.
CrO$_2$ is a half-metallic compound.
Sperlich $et$ $al$.~\cite{Sperlich13} reported Cr 2$p$ HAXPES of CrO$_2$, revealing a characteristic double-peak structure in the Cr 2$p_{3/2}$ main line. 
The low-$E_B$ feature in the double-peak structure is suppressed in a surface-sensitive soft x-ray spectrum, suggesting that a metallic CT screening is active in the 2$p$ excitation.
In this study, we conduct a HAXPES measurement of the Cr 1$s$ core-level in CrO$_2$, and compare the 1$s$ and 2$p$ spectra. Though both oxides are textbook examples of correlated metals, recent advances in, e.g.,~heterostructure and thin-film engineering have promoted renewed interest in their electronic structure and core-level HAXPES spectra~\cite{Mondal23,Aetukuri13,Shiga23,Fujiwara18}.

We simulate the experimental spectra using the Anderson impurity model (AIM) based on the local density approximation (LDA) + dynamical mean-field theory (DMFT) and conventional MO$_6$ cluster model. 
The LDA+DMFT method can be viewed as an extension of the MO$_6$ cluster model for computing the core-level XPS spectra. In this approach, the discrete ligand orbitals of the cluster model are replaced by a continuous metallic bath optimized within the LDA+DMFT scheme, which contains information about the entire lattice.
Thus, the LDA+DMFT AIM represents the CT screening responses, including the nonlocal CT channels~\cite{hariki17,ghiasi2019,higashi22}. The cluster model considers only the CT screening from the nearest-neighboring oxygens, traditionally called local screening~\cite{groot_kotani}. Thus, a comparison of the two models enables us to assess the contributions of the metallic nonlocal screening in the 1$s$ and 2$p$ spectra. The impurity model description allows us to consider the local core-valence multiplet interaction explicitly.

%%%%%%%%%%%%%%%%%%%%%%%%%%%%%%%%%%%%%%%%%%%%%%%%%%%%%%%%%
\section{Methods}
%%%%%%%%%%%%%%%%%%%%%%%%%%%%%%%%%%%%%%%%%%%%%%%%%%%%%%%%%

The Cr 1$s$ HAXPES of CrO$_2$ was measured at the HIKE station of the KMC-1 beamline at BESSY. A photon energy of 8.95~keV was used, and the energy resolution is $\sim$0.5~eV~\cite{ghiasi2019}. The CrO$_2$ sample was a commercially available crystalline CrO$_2$ powder (Magtrieve$^{\rm TM}$)~\cite{Sigma_Aldrich}. The powder was ground using a mortar and pressed into a pill. The V 1$s$ HAXPES spectrum of VO$_2$ was reproduced from the previous study~\cite{eguchi08}.

%\textcolor{red}{Anna: It would be great to say something about the monochromator crystal(s) used for the measurement and how the energy resolution was determined. In addition, information on the analyser would be good to include. Also - is there a beamline paper for the beamline we could cite?}

%\textcolor{red}{Anna: Re sample prep - was it an agate mortar? And was it a pill or a pellet? And any idea about the pressure used to prepare it?}

\begin{comment}
We use a 
commercially available 
crystalline CrO$_2$ powder (MAGTRIEVE$^{\rm TM}$)
with a density of $\rho=4.85~\rm g /cm^{3}$ at $25^\circ$C and grain size of 44~$\mu$m~\cite{Sigma_Aldrich}.
~(mesh 325)
The powder was milled by hand with mortar and pestle and then pressed into a pill. (This paragraph is taken from p.136 in Zimmermann's thesis: https://dspace.library.uu.nl/handle/1874/363530)
\end{comment}

%\textcolor{red}{Experimental detais ofs Cr 1$s$ HAXPES ...}

The simulation of Cr (V) 1$s$ and 2$p$ HAXPES for CrO$_2$ (VO$_2$) starts with a standard LDA+DMFT calculation~\cite{georges96,kotliar06,kunes09}. 
\textcolor{black}{Both CrO$_2$ and VO$_2$ (metallic phase) crystallize in a rutile structure with space group $P4_2/mnm$ (No.~136). The experimental lattice parameters of $a = 4.421$~\AA~(4.556~\AA) and $c=2.916$~\AA~(2.859~\AA) in CrO$_2$ (VO$_2$)~\cite{Anwar13,mcwhan74} are adopted in this work.} 
We perform an LDA calculation for the experimental crystal structure and subsequently construct a tight-binding model spanning both Cr (V) 3$d$ and O 2$p$ states from the LDA bands using Wien2K and wannier90 packages~\cite{wien2k,wien2wannier,wannier90}.
%For VO$_2$, a high-temperature rutile structure was used to study the metallic phase. 
%The LDA bands obtained for the experimental structures~\textcolor{black}{[ref]} are mapped onto a tight-binding model spanning Cr(V)~3$d$ and O~2$p$ orbitals~\cite{wien2k,wien2wannier,wannier90}.
The tight-binding model is augmented with local electron-electron interaction within the TM $3d$ shell that is determined by Hubbard $U$ and Hund's $J$ parameters. We set these values to $(U,J)$=(5.0~eV, 1.0~eV) and (6.0~eV, 1.0~eV) for the Cr and V cases, respectively, consulting with previous density functional theory (DFT)-based and spectroscopy studies for chromium and vanadium oxides~\cite{craco03,brito16}.
The strong-coupling continuous-time quantum Monte Carlo (CT-QMC) impurity solver~\cite{werner06,boehnke11,hafermann12,Hariki15} 
\textcolor{black}{with density-density approximation to the on-site electron-electron interaction}
is used to obtain the self-energies $\Sigma(i\omega_n)$ of Cr (V) $3d$ electrons from the AIM. The double-counting correction $\mu_{\rm dc}$, which is introduced to subtract the $d$--$d$ interaction effect present already in LDA step~\cite{karolak10,kotliar06}, is treated as an adjustable parameter and fixed to reproduce the experimental valence and core-level spectra.
The $\mu_{\rm dc}$ dependence of the calculated spectra can be found in the Supplementary Material (SM)~\cite{sm}.
To compute valence spectra and hybridization densities $\Delta(\varepsilon)$ on the real-frequency axis, analytically continued $\Sigma(\varepsilon)$ in the real-frequency domain is obtained using the maximum entropy method~\cite{jarrell96}. 
\textcolor{black}{We do not implement the charge self-consistency (updating the DFT charge) in the present LDA+DMFT calculation. Experimentally, VO$_2$ is a paramagnetic metal in a high-temperature phase~\cite{Morin59} and CrO$_2$ is a ferromagnetic metal with a Curie temperature of approximately 390~K~\cite{Swoboda61}. Thus, the calculations are performed for the paramagnetic phase for VO$_2$ and the ferromagnetic phase for CrO$_2$. }

The Cr (V) 1$s$ and 2$p$ HAXPES spectra are computed from the AIM implementing the DMFT hybridization densities $\Delta(\varepsilon)$, following methods detailed in Refs.~\cite{hariki17,ghiasi2019,winder20}.
\textcolor{black}{In this step, the AIM is extended to include the core orbitals and their interaction with the valence electrons.}
\textcolor{black}{Though the core-valence interaction value $U_{dc}$ is an adjustable parameter in this approach, an empirical relation $U_{dc}=1.2 \sim 1.3 \times U_{dd}$ is well established for 3$d$ TMOs, where $U_{dd}$ represents the configuration-averaged interaction on the TM 3$d$ shell~\cite{ghiasi2019,hariki17}. In SM~\cite{sm}, we verified that varying $U_{dc}$ within this range does not affect the spectra quantitatively, and thus it is set to $U_{dc} = 1.2 \times U_{dd}$ in the main result.}
%Coulomb interaction between the Cr (V) 2$p$ core-hole and 3$d$ electrons $U_{dc}$ 
%, where $U_{dd}$ represents the configuration-averaged interaction on the 3$d$ shell~\cite{ghiasi2019}.
%and the value is $U_{dd}$=4.556~eV and 5.556~eV for Cr and V, respectively,~\cite{Ghiasi2019}.
The multipole term in the core-valence interaction is parameterized by the higher-order Slater integrals obtained from the atomic Hartree-Fock code. 
\textcolor{black}{The atomic Slater integral values are reduced to 80~\% from the bare values to simulate the effect of intraatomic configuration interaction from higher basis configurations neglected in the atomic calculation. This treatment was successfully applied in modelling XPS as well as core-level x-ray absorption spectra in a wide range of 3$d$ TM compounds~\cite{Groot90,matsubara05,sugar72,tanaka92}. For CrO$_2$, the simulations implementing different scaling factors ($R_{2p-3d}$) to the bare values are performed.}
\textcolor{black}{The small spin-orbit coupling within the TM 3$d$ shell ($\xi_{\rm soc}\sim 0.04$~eV) is ignored in the LDA+DMFT self-consistent calculations to avoid complex off-diagonal elements in $\Delta(\varepsilon)$ in the CT-QMC calculation for the AIM, while it is considered explicitly when computing the core-level HAXPES spectra. The spin-orbit coupling in the TM 2$p$ shell  is also considered, and the spin-orbit coupling constants are taken from the atomic calculation~\cite{ghiasi2019,hariki17}.}

We employ the configuration-interaction (CI) solver for the AIM with the same implementation as described in Refs.~\cite{hariki17, winder20, hariki20}.
In the MO$_6$ cluster-model calculation, the electron hopping amplitudes with the nearest-neighboring ligands are extracted from the tight-binding model constructed from the LDA bands mentioned above and summarized in Appendix~\ref{appendix:cluster_param}.
The cluster model implements the same local Hamiltonian for the x-ray excited Cr (V) ion with the LDA+DMFT AIM.
As noted in Refs.~\cite{Hariki22,winder20}, care must be taken for the number of electronic configurations included to represent the XPS initial and final states within the CI scheme for a highly covalent system. We checked the convergence of the XPS spectral intensities as a function of the number of configurations in the studied compounds in Appendix~\ref{appendix:cluster_param}.

%%%%%%%%%%%%%%%%%%%%%%%%%%%%%%%%%%%%%%%%%%%%%%%%%%%%%%%%%
\section{Results}
%%%%%%%%%%%%%%%%%%%%%%%%%%%%%%%%%%%%%%%%%%%%%%%%%%%%%%%%%

\begin{comment}

We start our investigation with metallic VO$_2$. Figure~\ref{VO2_dos} shows valence-band spectra calculated by the LDA+DMFT method, together with the experimental valence-band XPS spectrum taken from Ref.~\cite{eguchi08}. As previously discussed in Ref.~\cite{biermann05}, while the single-site DMFT lacks the V--V intersite self-energy needed for describing the dimerized insulating phase with a monoclinic structure, it provides a reasonable description for the correlated metallic phase of the high-temperature rutile-structure phase. Our DMFT valence-band spectrum is similar with the cluster DMFT result for the metallic phase~\cite{brito16,biermann05}. The LDA+DMFT spectrum exhibits an incoherent lower Hubbard band feature around $-1$~eV and a coherent metallic feature near the Fermi energy $E_F$. The calculated binding energies $E_B$ of the O 2$p$ states with respect to the V 3$d$ states match well with the experimental data, supporting the used value of the double-counting correction $\mu_{\rm dc}$ in the LDA+DMFT calculation.

\end{comment}
 
%%%%%%%%%%%%%%%%%%%%%%%%%%%%%%%%%%%%%%%%%%%%%%%%%%%%%%%%%
\begin{figure}[t]
    \includegraphics[width=0.99\columnwidth]{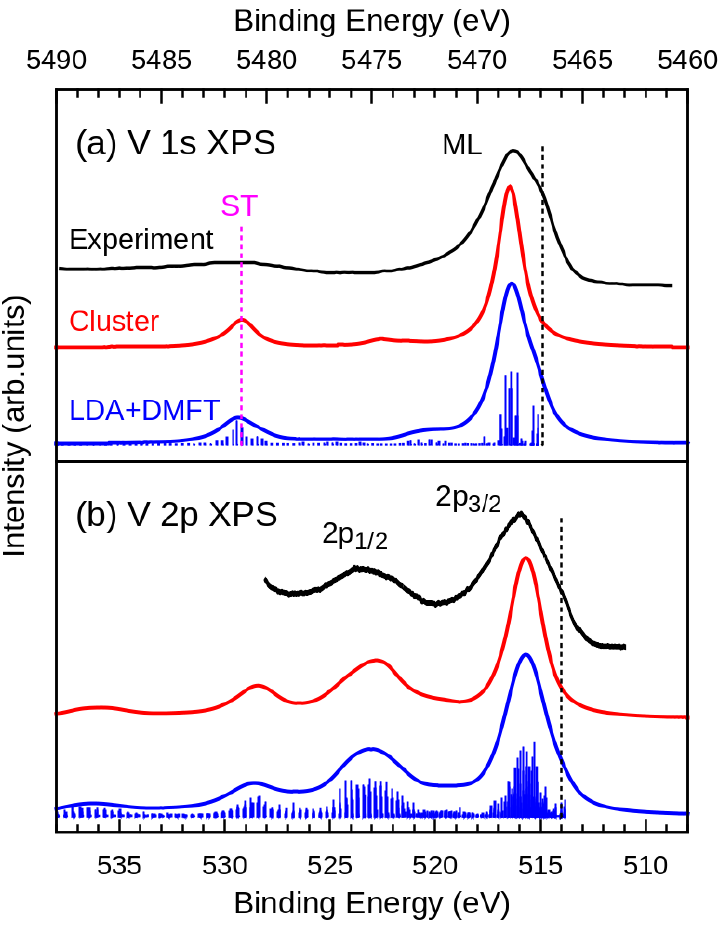}
    \caption{The LDA+DMFT AIM (blue) and the VO$_6$ cluster model (red) calculations for (a) V 1$s$ XPS and (b) V 2$p$ XPS spectra for metallic VO$_2$ in the high-temperature rutile structure phase. The experimental HAXPES data are taken from Ref.~\cite{eguchi08}. The spectral intensities are convoluted with a Gaussian of 200~meV and a Lorentzian of 650~meV in the half width at half maximum (HWHM) for both 1$s$ and 2$p$ spectra. The main line (ML) and the satellite (ST) are indicated.
    %\textcolor{red}{Anna: I think it would be good to also add Binding Energy (eV) to the top x axis}
    }
    \label{VO2_xps}
\end{figure}
%%%%%%%%%%%%%%%%%%%%%%%%%%%%%%%%%%%%%%%%%%%%%%%%%%%%%%%%%

Figure~\ref{VO2_xps} shows the V 1$s$ and 2$p$ XPS spectra of metallic VO$_2$ calculated by the LDA+DMFT AIM method and the VO$_6$ cluster model.
The experimental 1$s$ and 2$p$ XPS data are reproduced from Ref.~\cite{eguchi08}.
The LDA+DMFT valence-band spectra can be found in Appendix~\ref{appendix:valence_spectra}.
Apart from the presence of the V 2$p_{1/2}$ spin-orbit component and a different lifetime broadening, the V 1$s$ and 2$p_{3/2}$ experimental spectra exhibit similar line shapes. 
A satellite (ST) feature is located approximately 13~eV above the main line (ML), as indicated by a dashed line.

\begin{figure}[t]
    \includegraphics[width=0.99\columnwidth]{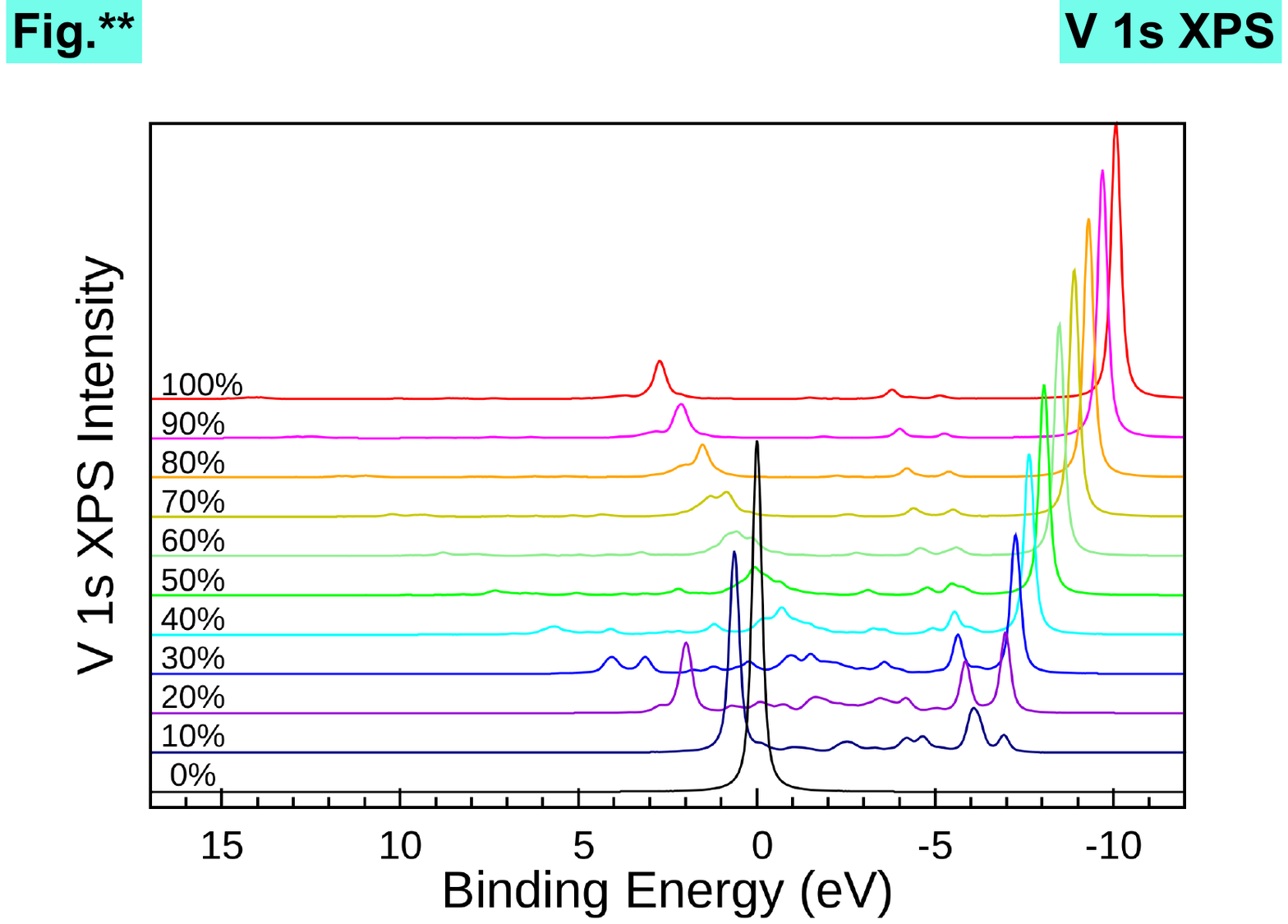}
    \caption{V 1$s$ XPS spectra simulated by the VO$_6$ cluster model with varying metal-ligand (V 3$d$--O 2$p$) hybridization strength. 100~\% corresponds to the hybridization strength derived from the LDA bands for the experimental rutile structure of VO$_2$ in Appendix~\ref{appendix:cluster_param}.
    %The diagonal energies of the electronic configuration for the 1$s$ XPS final states are indicated with the vertical bars.
    The spectral intensities are convoluted with a Gaussian of 100~meV and a Lorentzian of 100~meV (HWHM).}
    \label{fig_hyb_scan_v}
\end{figure}

The VO$_6$ cluster model represents the overall features in the V 1$s$ and 2$p$ spectra.
In Fig.~\ref{fig_hyb_scan_v}, the 1$s$ spectra computed with varying the metal-ligand hybridization in the cluster model are shown. The diagonal energies of the electronic configurations in the final states are listed in Table~\ref{Tab_vo2_fene};~see Appendix~\ref{appendix:cluster_param} for those in the initial state.
%, where $\underline{c}$ denotes a hole on the 1$s$ core level. 
Note that the screened $|\underline{c} d^2\underline{L}\rangle$ configuration has an energy lower than that of $|\underline{c} d^1\rangle$ due to the core-valence interaction $U_{dc}$ for an excess $d$ electron. 
%Configuration energies for the initial state are given in %Table~\ref{Tab_cluster_param} of Appendix~\ref{appendix:cluster_param}.
The effective hybridization $V_{\rm eff}$ with the ligands can be given as $V_{\rm eff}=[(4-N_{e_g})\times V^2_{e_g}+(6-N_{t_{2g}})\times V^2_{t_{2g}}]^{1/2}$, where $N_{e_g}$ ($N_{t_{2g}}$) and $V_{e_g}$ ($V_{t_{2g}}$) denote occupation of the $e_g$ ($t_{2g}$) states and hybridization strength of the ligand and $e_g$ ($t_{2g}$) orbitals~\cite{groot_kotani,ghiasi2019,Hariki22,uozumi97}.
For VO$_2$ (with $N_{t_{2g}}=1$ and $N_{e_{g}}=0$) in a formal valence state, $V_{\rm eff}$ accounts for 9.1~eV (Appendix~\ref{appendix:cluster_param}). 
The large hybridization leads to well-split bondinglike and anti-bondinglike states of the $|\underline{c}d^1\rangle$ and $|\underline{c}d^2\underline{L}\rangle$ states, forming the main line and satellite, respectively.
%The large $V_{\rm eff}$ results in strong mixed character between the $|\underline{c}d^1\rangle$ and $|\underline{c}d^2\underline{L}\rangle$ configurations. %, see the inset of Fig.~\ref{fig_hyb_scan_v} for the configuration weights. 
Similar to Ti oxides~\cite{Hariki22}, higher configurations up to $|\underline{c}d^5\underline{L}^4\rangle$ are necessary for accurate description of the satellites, as demonstrated in Appendix~\ref{appendix:cluster_param}.
%, similar to those in the Ti 1$s$ XPS spectra as pointed out in recent study~\cite{Hariki22}.

%%%%%%%%%%%%%%%%%%%%%%%%%%%%%%%%%%%%%%%%%%%%%%%%%%%%%%%%%
%%%%%%%%%%%%%%%%%%%%%%%%%%%%%%%%%%%%%%%%%%%%%%%%%%%%%%%%%

%%%%%%%%%%%%%%%%%%%%%%%%%%%%%%%%%%%%%%%%%%%%%%%%%%%%%%%%%%%%%%%
      \begin{table}
            \centering
            \begin{ruledtabular}            
            \begin{tabular}{ c  c  c  c  c  c  c  c  c  c  c }
      Conf.                                  & Diagonal energies                            & Value  \\
     \hline
     {$|{\underline{c}}d^{1}{\rangle}$}                     & 0                                  &  0.0  \\
     {$|{\underline{c}}d^{2}{\underline{L}}^{1}{\rangle}$}  &  {$\Delta_{\rm CT}-U_{dc}$}                &  $-4.7$  \\
     {$|{\underline{c}}d^{3}{\underline{L}}^{2}{\rangle}$}  & 2{$\Delta_{\rm CT}$}+{$U_{dd}-2U_{dc}$}    &  $-3.8$  \\
     {$|{\underline{c}}d^{4}{\underline{L}}^{3}{\rangle}$}  & 3{$\Delta_{\rm CT}$}+3{$U_{dd}-3U_{dc}$}   &   2.7  \\
     {$|{\underline{c}}d^{5}{\underline{L}}^{4}{\rangle}$}  & 4{$\Delta_{\rm CT}$}+6{$U_{dd}-4U_{dc}$}   &  14.7  \\
     {$|{\underline{c}}d^{6}{\underline{L}}^{5}{\rangle}$}  & 5{$\Delta_{\rm CT}$}+10{$U_{dd}-5U_{dc}$}  &  32.3  \\
     {$|{\underline{c}}d^{7}{\underline{L}}^{6}{\rangle}$}  & 6{$\Delta_{\rm CT}$}+15{$U_{dd}-6U_{dc}$}  &  55.4  \\
     \end{tabular}
\caption{The configuration (diagonal) energies in the V 1$s$ XPS final states within the VO$_6$ cluster-model Hamiltonian in eV.
\textcolor{black}{Here, the charge-transfer energy is defined for a V$^{4+}$ ($d^1$) electronic configuration as $\Delta_{\rm CT}=E(d^2\underline{L})-E(d^1)=\varepsilon_d-\varepsilon_L+U_{dd}$, where $\varepsilon_d$ ($\varepsilon_L$) denotes the averaged V $3d$ (ligand 2$p$) energy.}
The values of charge-transfer energy $\Delta_{\rm CT}$, 3$d$--3$d$ interaction $U_{dd}$ and 2$p$--3$d$ core-valence interaction $U_{dc}$ are found in Appendix~\ref{appendix:cluster_param}.}
\label{Tab_vo2_fene}
\end{ruledtabular}
\end{table}
%%%%%%%%%%%%%%%%%%%%%%%%%%%%%%%%%%%%%%%%%%%%%%%%%%%%%%%%%%%%%%%

Both the V 1$s$ and 2$p_{3/2}$ main lines exhibit a shoulder feature originating from metallic nonlocal CT screening at $E_B=5467$~eV in 1$s$ and $E_B=514$~eV in 2$p_{3/2}$, as indicated by the dotted lines in Fig.~\ref{VO2_xps}.
The LDA+DMFT AIM reproduces the shoulder feature, \textcolor{black}{while it is missing in the cluster-model result, as expected given that the former model includes a metallic CT screening, while the latter model considers only the metal-ligand one.
Furthermore, the LDA+DMFT AIM yields a broad band feature around $E_B=5472$~eV in the 1$s$ spectrum, which
%This feature %primarily consists of non-bonding-like final states and 
shares similarities with the low-$E_B$ satellites 
%reported recently
in Ti 1$s$ HAXPES spectra of TiO$_2$ and SrTiO$_3$~\cite{Woicik20,Woicik15,Hariki22}.} Due to its overlap with the broad tail of the V 2$p_{1/2}$, this feature is not clearly seen in the V 2$p_{3/2}$ spectrum [Fig.~\ref{VO2_xps}b]. 
%The VO$_6$ single cluster-model result lacks both the nonlocal-screening feature and the broad non-bonding feature. %, see Fig.~\ref{VO2_xps} and Fig.~\ref{fig_hyb_scan_v}.
%The VO$_6$ cluster-model result captures the overall line shape, while 
%it fails in simulating the shoulder feature due to the nonlocal screening in the main line nor the broad band shape of the non-bonding feature due to the lack of the band effect in the single VO$_6$ cluster description. 

%%%%%%%%%%%%%%%%%%%%%%%%%%%%%%%%%%%%%%%%%%%%%%%%%%%%%%%%%
\begin{figure}[t]
    \includegraphics[width=0.99\columnwidth]{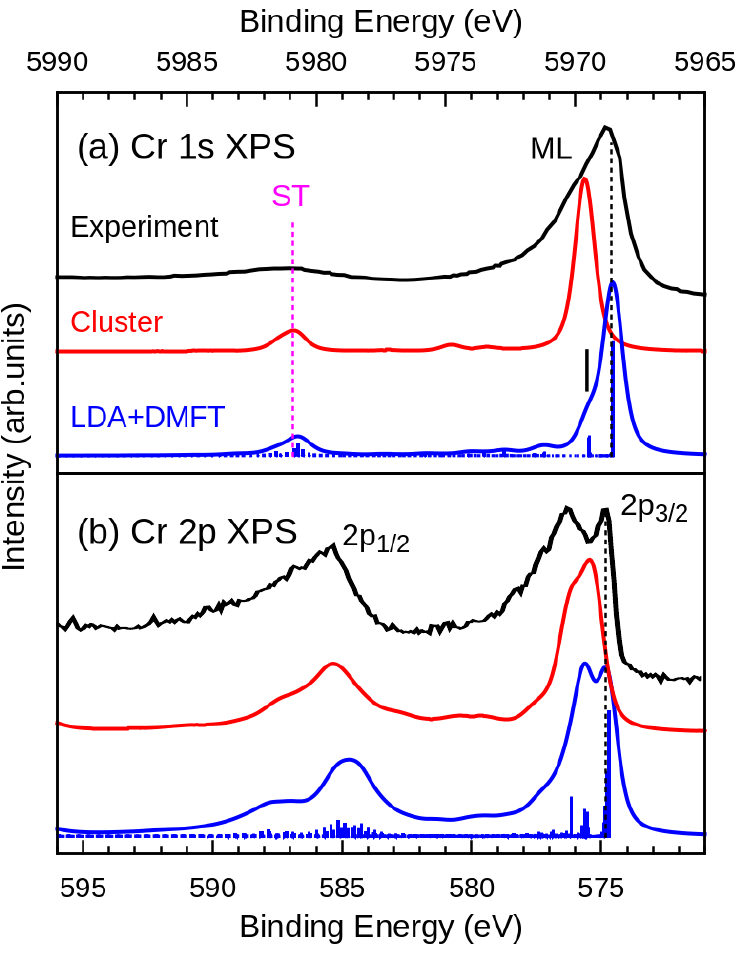}
    \caption{The experimental HAXPES, the LDA+DMFT AIM (blue), and the CrO$_6$ cluster-model (red) calculations for (a) Cr 1$s$ XPS and (b) Cr 2$p$ XPS spectra. The spectral intensities are convoluted with a Gaussian and a Lorentzian of 300 and 250~meV (HWHM), respectively, for both 1$s$ and 2$p$ spectra. The experimental Cr 2$p$ HAXPES data are taken from Ref.~\cite{Sperlich13}. The main line (ML) and the satellite (ST) are indicated.}
    \label{CrO2_xps}
\end{figure}
%%%%%%%%%%%%%%%%%%%%%%%%%%%%%%%%%%%%%%%%%%%%%%%%%%%%%%%%%

Next, we examine \textcolor{black}{ferromagnetic} metallic CrO$_2$.
%
%
%
%
%
%Figure~\ref{CrO2_dos} shows the LDA+DMFT valence-band spectral intensities, together with the experimental XPS data taken from Ref.~\cite{Sperlich13}.
%
%
%
%
%
%
%The double-counting dependence of the valence as well as the core-level spectra can be found in SM~\cite{sm}.
%
Figure~\ref{CrO2_xps} presents the Cr 1$s$ and 2$p$ XPS spectra calculated by the LDA+DMFT AIM and the CrO$_6$ cluster model.
The Cr 2$p$ experimental HAXPES spectrum is taken from Ref.~\cite{Sperlich13}.
The LDA+DMFT valence-band spectra can be found in Appendix~\ref{appendix:valence_spectra}.
The Cr 2$p_{3/2}$ main line ($E_B \sim 576$~eV) exhibits the double-peak feature.
%The intensity of the low-$E_B$ peak ($E_B \sim 575$~eV) is reduced in the surface-sensitive soft x-ray spectrum compared with the HAXPES spectrum~\cite{Sperlich13}, suggesting that the metallic screening effect is present in the spectra.
%\tc{blue}{\bf{[I suggest adding a few more lines about it. (by Masoud)]}}s
%The importance of metallic screening for the low-$E_B$ peak ($E_B \sim 575$~eV) for the formation of the double peaks was pointed out in Ref.~\cite{Sperlich13}.
Interestingly, the Cr 1$s$ main line ($E_B \sim 5968.5$~eV) shows an asymmetric line shape, but not a double-peak feature as in the Cr 2$p_{3/2}$ spectrum.
This is a notable contrast with VO$_2$ (Fig.~\ref{VO2_xps}), as the V 1$s$ and 2$p_{3/2}$ spectra resemble each other.
%Therefore, we anticipate that the Cr 1$s$ spectra provide qualitatively distinct information from the 2$p$ spectra.
The LDA+DMFT results reproduce both the Cr 1$s$ and 2$p$ spectra reasonably well. The cluster model incorrectly yields a symmetric single-peak 1$s$ main line and does not reproduce the double-peak feature in the Cr 2$p_{3/2}$ main line (Fig.~\ref{CrO2_xps}).
This suggests the presence of metallic CT contributions in both the Cr 1s and 2$p$ spectra of CrO$_2$ as in the VO$_2$ case. 
\textcolor{black}{A detailed analysis of the spectra within the CrO$_6$ cluster model can be found in Appendix~\ref{appendix:cluster_cro2}.}

%%%%%%%%%%%%%%%%%%%%%%%%%%%%%%%%%%%%%%%%%%%%%%%%%%%%%%%%%
\begin{figure}[t]
    \includegraphics[width=0.99\columnwidth]{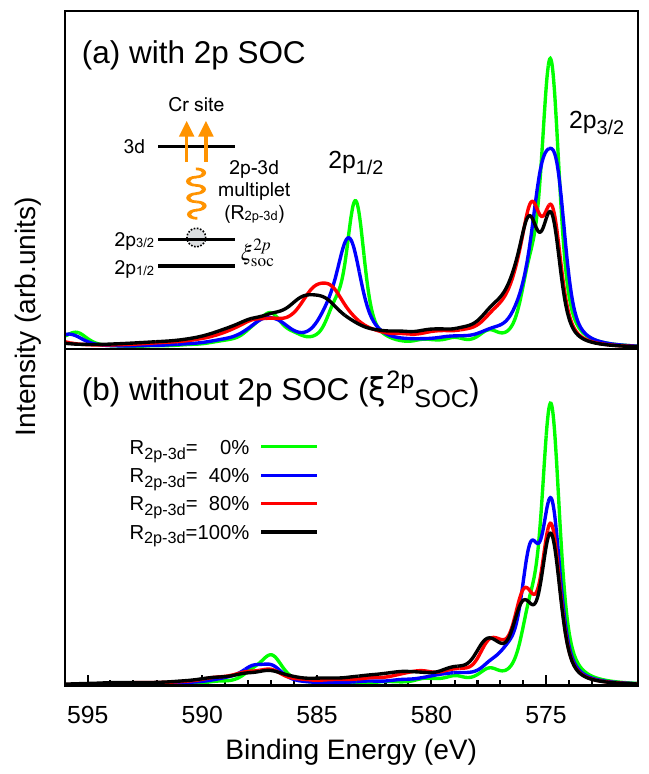}
    \caption{(a) Cr 2$p$ XPS spectral intensities calculated by the LDA+DMFT AIM method with reducing strength of the 2$p$--3$d$ core-valence Coulomb multiplet interaction from the ionic values. (b) The calculated spectra without the spin-orbit coupling ($\xi^{2p}_{\rm SOC}=0$) on the 2$p$ core orbitals.
    The spectral intensities are convoluted with a Gaussian and a Lorentzian of 300 and 250~meV (HWHM), respectively.}
    \label{CrO2_red}
\end{figure}
%%%%%%%%%%%%%%%%%%%%%%%%%%%%%%%%%%%%%%%%%%%%%%%%%%%%%%%%%

To get insight into the different shapes in the 1$s$ and 2$p$ main line in CrO$_2$, we calculate the 2$p$ spectra with modulating the multiplet coupling between the Cr 2$p$ and 3$d$ electrons in Fig.~\ref{CrO2_red}(a). 
Here, $R_{2p-3d}$ denotes a scaling factor for the multipole part in the Cr 2$p$--3$d$ interaction in the LDA+DMFT AIM. 
The 2$p$--3$d$ core-valence multiplet interaction significantly influences the 2$p_{3/2}$ main-line shape. 
The 2$p$ spectrum with $R_{2p-3d}=0\%$ closely resembles the Cr 1$s$ spectrum in Fig.~\ref{CrO2_xps}(a).
This indicates that the CT screening to the x-ray excited Cr from the rest of the crystal is essentially the same in the 1$s$ and 2$p$ excitations, which is not surprising, given that the hybridization strength of the Cr $3d$ states with the electron bath is determined in the valence electronic structure when a core hole is absent. However, the 2$p$--3$d$ multiplet interaction at the x-ray excited Cr site mixes the screened states via different CT channels substantially, and consequently the intrinsic CT spectral peaks are distributed in a complex way in the 2$p$ spectrum in the case of CrO$_2$.  The 1$s$ excitation, with negligibly weak core-valence interaction, is thus better suited for examining and quantifying the CT effect.

\textcolor{black}{In the 1$s$ spectrum, the prominent peak at the low-$E_B$ side ($\sim 5968.5$~eV, dashed line) dominantly consists of the metallic-screening states, while the high-$E_B$ feature ($\sim 5969.5$~eV, solid bar) contains mostly the local-screening states from ligands. Only the latter CT channel can be captured in the CrO$_6$ cluster model [see Fig.~\ref{CrO2_xps}(a)]. The LDA+DMFT AIM result indicates that the intrinsic spectral intensity of the metallic screening is substantial compared to the local screening one upon the core-level excitation, 
manifesting in the observed asymmetric line shape in the experimental 1$s$ data. However, the intensities of the two CT features are largely modulated due to the strong 2$p$--3$d$ multiplet interaction in the 2$p$ spectrum. Thus, the double peaks in the 2$p_{3/2}$ spectrum do not serve as a good measure of the two CT channels.}
%
%
%the enhanced peak at the low-$E_B$ side ($E_B \sim 575$~eV in 2$p_{3/2}$ and $\sim 5968.5$~eV in 1$s$) indicated by a dashed line consists of nonlocal-screened final states dominantly, while the high-$E_B$ feature ($\sim 576$~eV in 2$p_{3/2}$ and $\sim 5969.5$~eV in 1$s$) contains mostly the local-screened states.
%
%The 2$p$--3$d$ core-valence multiplet interaction
%substantially mixes 
%acts on the two screened final states differently, resulting in the double-peak shape of the Cr 2$p_{3/2}$ HAXPES main line. 
%This results in the double-peak structure in the Cr 2$p_{3/2}$ main line.
%Thus, the Cr 1$s$ XPS spectrum with negligibly weak core-valence interaction effect is better suited than the 2$p$ XPS for examining and quantifying the charge transfer effect and chemical-bonding property. 
%Namely, the nonlocal screening feature in the low-$E_B$ component ($\sim 5968.5$~eV) of the 1$s$ main line has much larger spectral intensity than the local screening feature in the high-$E_B$ component ($\sim 5969.5$~eV).
%As demonstrated in Fig.~\ref{CrO2_red}(a), however the 2$p$--3$d$ core-valence Coulomb multiplet terms mixes the locally- and nonlocally-screened final-state wave functions, leading to a large spectral redistribution between the two features. 

Figure~\ref{CrO2_red}(b) shows the 2$p$ XPS spectra calculated with eliminating the spin-orbit interaction on the Cr 2$p$ shell in the LDA+DMFT AIM Hamiltonian. The spin-orbit interaction plays a marginal role for the 2$p_{3/2}$ main-line shape and does not affect the spectra in the absence of the 2$p$--3$d$ interaction ($R_{2p-3d}=0\%$). Thus, the 2$p$--3$d$ core-valence multiplet interaction is the decisive factor for the difference in the Cr 1$s$ and 2$p$ spectra.

\begin{comment}

Finally, we briefly discuss the interplay between the formation of the atomic multiplet and Cr--O hybridization in the Cr core-level spectra using the CrO$_6$ cluster model. 
%Although the cluster model lacks the nonlocal screening effect, its simpler excitation spectrum 
%compared to the LDA+DMFT AIM one with the continuous hybridization density 
%allows us to examine the influence of the Cr--O bonding effect.
The configuration energies in the XPS final states within the cluster-model Hamiltonian are summarized in Table~\ref{Tab_cro2_fene}. 
In Fig.~\ref{fig_hyb_scan_cr}, we calculate the Cr 1$s$ and 2$p$ XPS spectra with changing the Cr--O orbital hybridization strength. In the atomic limit (0~\%), i.e.,~$V=0$, the Cr 2$p$ XPS shows richer multiplet peaks in both the 2$p_{3/2}$ and 2$p_{1/2}$ components due to the strong Cr 2$p$--3$d$ multipole interaction. 
%Note that due to the negative charge-transfer energy ($\Delta_{\rm CT}=-0.5$~eV), the ground state takes a locally $d^3$ configuration. 
The Cr--O hybridization renormalizes substantially the multiplet peaks, Fig.~\ref{CrO2_xps}b. The negligibly-weak multiplet effect on the 1$s$ excitation yields a quasi single-peak line in the atomic limit (Fig.~\ref{CrO2_xps}a).

\end{comment}

%%%%%%%%%%%%%%%%%%%%%%%%%%%%%%%%%%%%%%%%%%%%%%%%%%%%%%%%%
\section{Concluding Remarks}
%%%%%%%%%%%%%%%%%%%%%%%%%%%%%%%%%%%%%%%%%%%%%%%%%%%%%%%%%

In this study, we have analyzed 1$s$ and 2$p$ core-level HAXPES spectra of metallic VO$_2$ and CrO$_2$, archetypal correlated metallic 3$d$ transition-metal oxides. We conducted a HAXPES measurement for the 1$s$ core level of CrO$_2$.
While the V 1$s$ spectrum in VO$_2$ closely resembles the 2$p_{3/2}$ spectrum in the literature, we find the Cr 1$s$ main line in CrO$_2$, exhibiting asymmetric line shape, substantially differs from the Cr 2$p_{3/2}$ main line with the double peaks.
Through a comprehensive computational analysis for the 1$s$ and 2$p$ spectra of VO$_2$ and CrO$_2$
using the LDA+DMFT AIM method and conventional MO$_6$ cluster model, we demonstrated that the difference in the CrO$_2$ and VO$_2$ spectra originates from a strong core-valence atomic multiplet effect in the 2$p$ excitation of CrO$_2$. The double peaks in the Cr 2$p_{3/2}$ main line do not directly represent either of the metallic-screening or metal-ligand CT final states, as these CT final states are mixed by the multiplet interaction acting locally at the x-ray excited Cr site. 
The core-valence multiplet-free 1$s$ spectrum unveils intrinsic spectral features of the metallic and metal-ligand CT screenings upon the core-level excitation. Our LDA+DMFT AIM calculation for the 1$s$ spectra reveals that the spectral intensity of the metallic screening dominates that of the metal-ligand screening, highlighting the advantage of 1$s$ excitation in characterizing correlated metallic 3$d$ TMOs,
%over the routinely-employed 2$p$ excitation 
especially when the atomic multiplet effect is not negligible in the 2$p$ spectra. Future systematic investigations of the 1$s$ HAXPES spectra for 3$d$ TMOs with the active 2$p$--3$d$ multiplets, such as Co or Mn oxides, would be of interest.

%%%%%%%%%%%%%%%%%%%%%%%%%%%%%%%%%%%%%%%%%%%%%
%%%%%%%%%%%%%%%%%%%%%%%%%%%%%%%%%%%%%%%%%%%%%
%%%%%%%%%%%%%%%%%%%%%%%%%%%%%%%%%%%%%%%%%%%%%

\begin{acknowledgments}

A.H. thank Hidenori Fujiwara for fruitful discussions. A.H. was supported by JSPS KAKENHI with Grant Numbers 21K13884, 21H01003, 23K03324, 23H03816, 23H03817. 
\end{acknowledgments}

%%%%%%%%%%%%%%%%%%%%%%%%%%%%%%%%%%%%%%%%%%%%%
%%%%%%%%%%%%%%%%%%%%%%%%%%%%%%%%%%%%%%%%%%%%%
%%%%%%%%%%%%%%%%%%%%%%%%%%%%%%%%%%%%%%%%%%%%%

\appendix

\section{LDA+DMFT valence spectra}
\label{appendix:valence_spectra}

\begin{figure}
    \includegraphics[width=0.99\columnwidth]{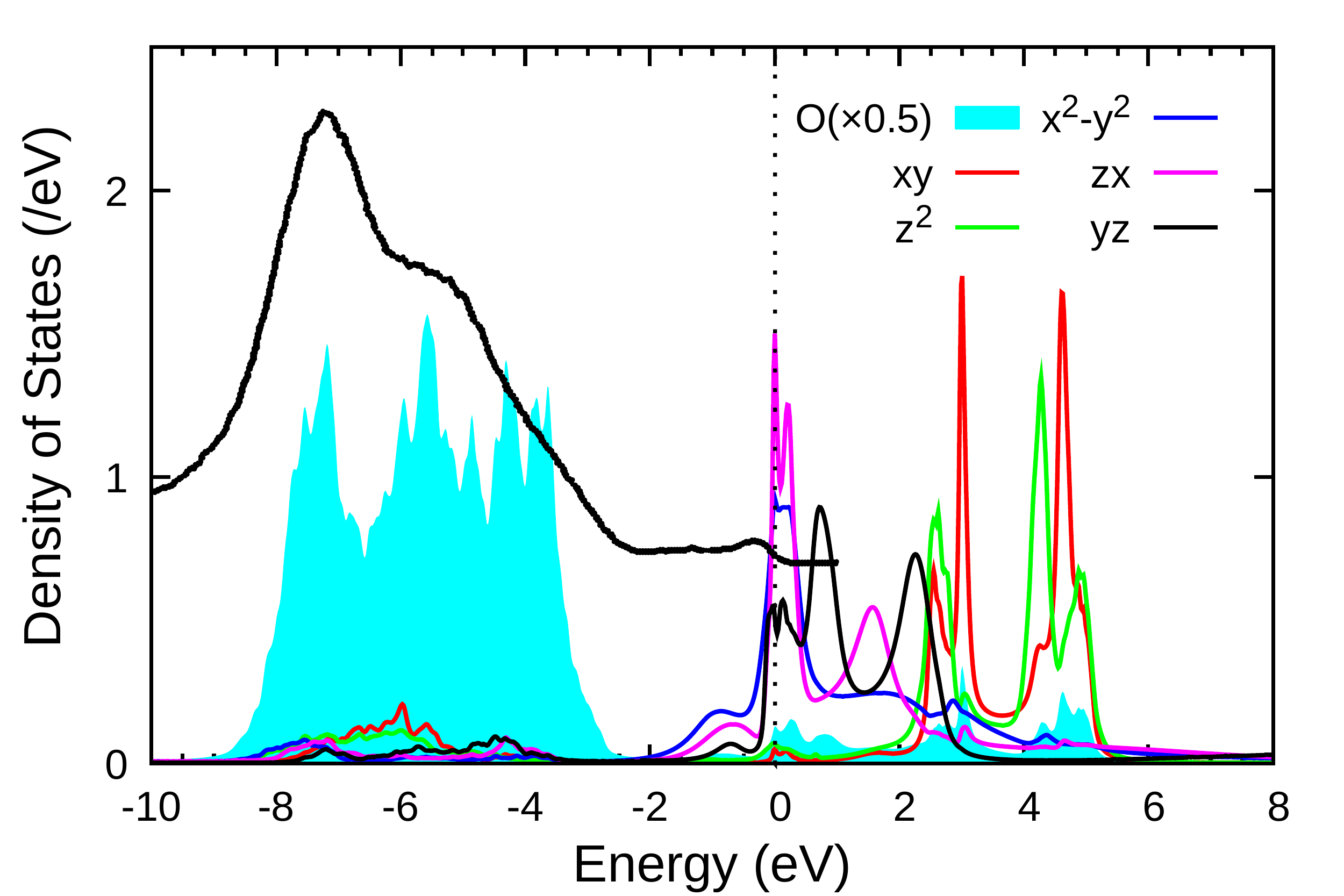}
    \caption{The LDA+DMFT valence spectral intensities for metallic bulk VO$_2$. The experimental valence-band XPS spectrum measured at 320~K is taken from Ref.~\cite{eguchi08}.}
    \label{VO2_dos}
\end{figure}

\begin{figure}[t]
    \includegraphics[width=0.99\columnwidth]{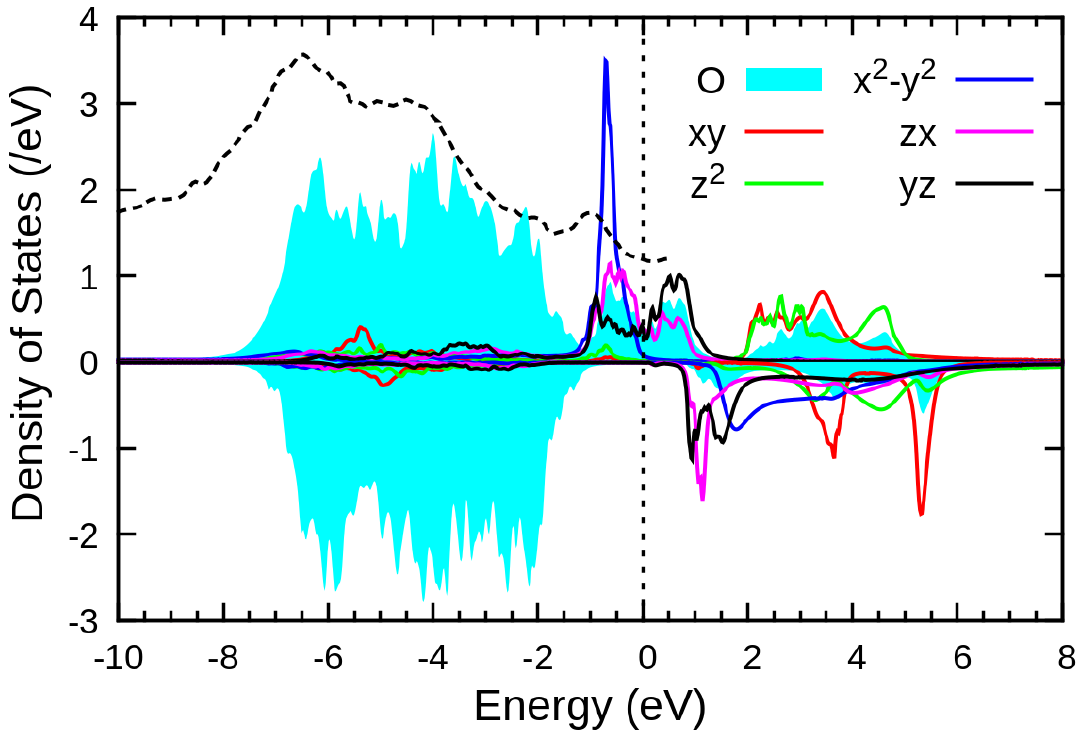}
    \caption{The LDA+DMFT valence spectral intensities for bulk CrO$_2$. The experimental valence-band XPS spectrum (dashed) measured at 150~K is taken from Ref.~\cite{Sperlich13}.}
    \label{CrO2_dos}
\end{figure}

Figure~\ref{VO2_dos} shows valence-band spectra of metallic VO$_2$ calculated by the LDA+DMFT method,
together with the experimental valence-band XPS spectrum taken from Ref.~\cite{eguchi08}. As previously discussed in Ref.~\cite{biermann05}, while the single-site DMFT lacks the V--V intersite self-energy needed for describing the dimerized insulating phase with a monoclinic structure, it provides a reasonable description for the correlated metallic phase of the high-temperature rutile-structure phase. Our DMFT valence-band spectrum is similar to the cluster DMFT result for the metallic phase~\cite{brito16,biermann05}. 
%\tc{blue}{\bf{[What do you mean by "incoherent" and "coherent"? (by Masoud)]}}
The LDA+DMFT spectrum exhibits a lower Hubbard-band feature around $-1$~eV and a quasiparticle feature near the Fermi energy $E_F$. The calculated binding energies $E_B$ of the O 2$p$ states with respect to the V 3$d$ states match well with the experimental data, supporting the used value of the double-counting correction $\mu_{\rm dc}$ in the LDA+DMFT calculation.
In Fig.~\ref{CrO2_dos}, the LDA+DMFT valence-band spectra of CrO$_2$ together with the experimental valence-band XPS spectrum taken from Ref.~\cite{Sperlich13} is also shown. The double-counting dependence of the valence spectra of VO$_2$ and CrO$_2$ can be found in the SM~\cite{sm}.

%%%%%%%%%%%%%%%%%%%%%%%%%%%%%%%%%%%%%%%%%%%%%%%%%%%%%%%%%
%%%%%%%%%%%%%%%%%%%%%%%%%%%%%%%%%%%%%%%%%%%%%%%%%%%%%%%%%

%%%%%%%%%%%%%%%%%%%%%%%%%%%%%%%%%%%%%%%%%%%%%%%%%%%%%%%%%
%%%%%%%%%%%%%%%%%%%%%%%%%%%%%%%%%%%%%%%%%%%%%%%%%%%%%%%%%

%%%%%%%%%%%%%%%%%%%%%%%%%%%%%%%%%%%%%%%%%%%%%%%%%%%%%%%%%%%%%%%
\begin{figure}
    \includegraphics[width=0.90\columnwidth]{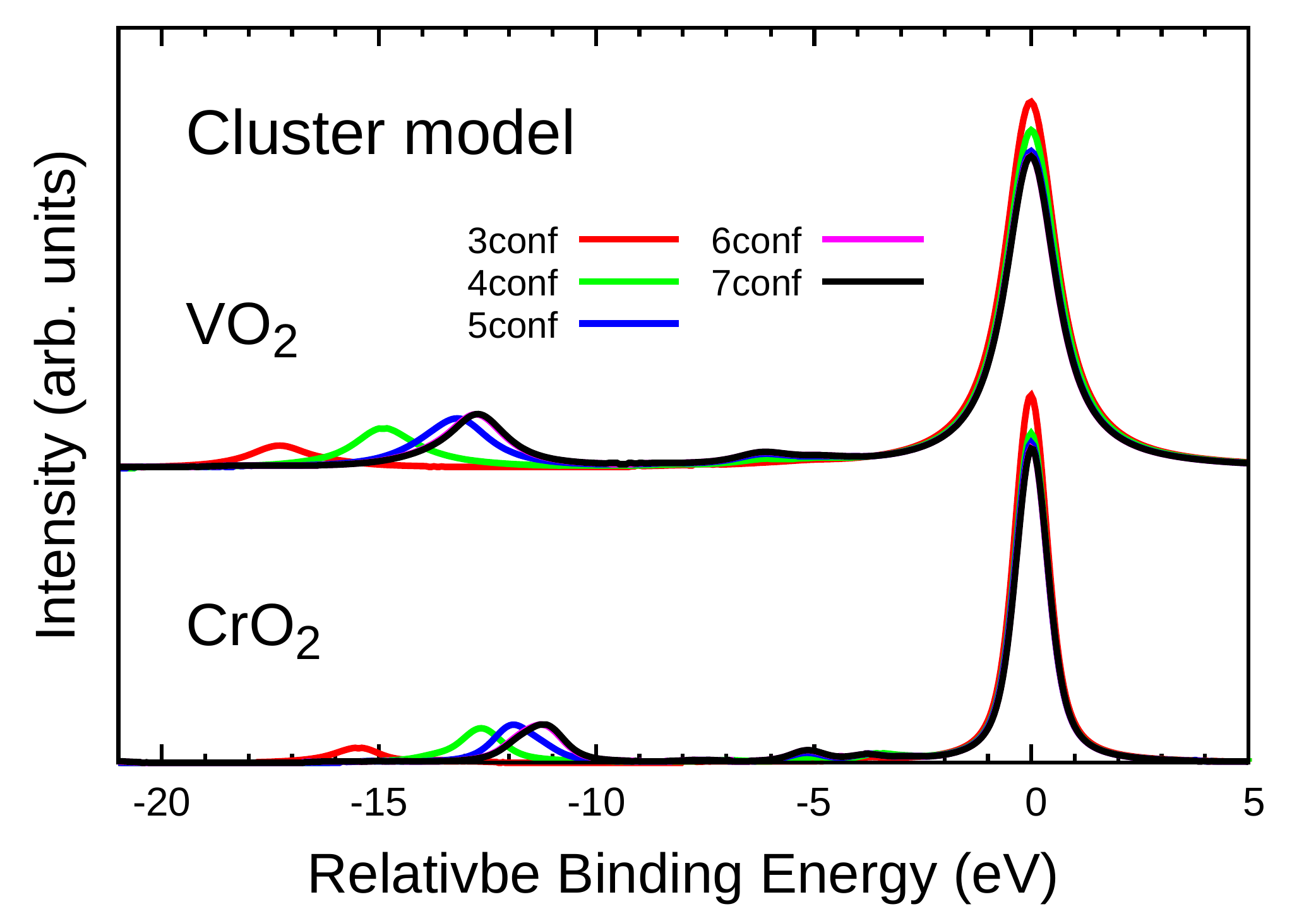}
    \caption{The V (Cr) 1$s$ XPS spectra calculated by the VO$_6$ (CrO$_6$) cluster model with varying the number of electronic configurations included in the spectral calculation. Here, the spectral intensities are convoluted with a Gaussian of 200 (300)~meV and a Lorenzian of 650 (250)~meV for VO$_2$ (CrO$_2$) (HWHM).}
    \label{fig_conf_dep}
\end{figure}
%%%%%%%%%%%%%%%%%%%%%%%%%%%%%%%%%%%%%%%%%%%%%%%%%%%%%%%%%%%%%%%

\section{Model parameters in the cluster model calculation}
\label{appendix:cluster_param}

%Table~\ref{Tab_ander_weight} shows the weights of the electronic configurations in the initial state for VO$_2$ calculated with increasing the number of the electronic configurations included in the LDA+DMFT AIM basis expansion. Here, $C$ denotes an electron in conduction band.

Table~\ref{Tab_cluster_param} summarizes the parameter values employed in our cluste- model calculations for VO$_2$ and CrO$_2$. 
The interaction parameters $U_{dd}$ and $U_{dc}$ are set to the same values as the ones in the LDA+DMFT AIM Hamiltonian for VO$_2$ and CrO$_2$. 
The one-particle parameters, i.e.,~electron hopping amplitudes and crystal-field energies, are derived from the tight-binding models constructed from the DFT bands for the experimental structures.

The hopping amplitudes between the V (Cr) $d$ orbitals [$B_{1g}(xy)$, $A_g(3z^2-r^2)$, $A'_g(x^2-y^2))$, $B_{2g}(zx)$ and $B_{3g}(yz)$] and the molecular orbitals consisting of the nearest-neighboring ligand 2$p$ orbitals with the same symmetry are given in Table~\ref{Tab_cluster_param}. The charge-transfer energy $\Delta_{\rm CT}$ is taken from Refs.~\cite{Uozumi93, Chang05}.
Table~\ref{Tab_cluster_iene} summarizes the diagonal energies of the electronic configurations considered in the VO$_6$ and CrO$_6$ cluster model accounting for the interaction $U_{dd}$ and the charge-transfer energy $\Delta_{\rm CT}$ up to the $|d^{7}{\underline{L}}^{6}{\rangle}$ configuration.
Figure~\ref{fig_conf_dep} shows the V and Cr 1$s$ XPS spectra calculated by the cluster model with increasing the number of the electronic configurations included in representing the initial and final states. 
The position of the satellites is converged well with the basis set including up to the $|d^5\underline{L}^4\rangle$ configuration for VO$_2$ and the $|d^6\underline{L}^4\rangle$ configuration for CrO$_2$.
We also checked the convergence of the LDA+DMFT AIM spectra with respect to the number of the electronic configurations included in the configuration-interaction solver evaluation~\cite{hariki17,winder20} of the spectral intensities.

%%%%%%%%%%%%%%%%%%%%%%%%%%%%%%%%%%%%%%%%%%%%%%%%%%%%%%%%%
\begin{figure}
    \includegraphics[width=0.99\columnwidth]{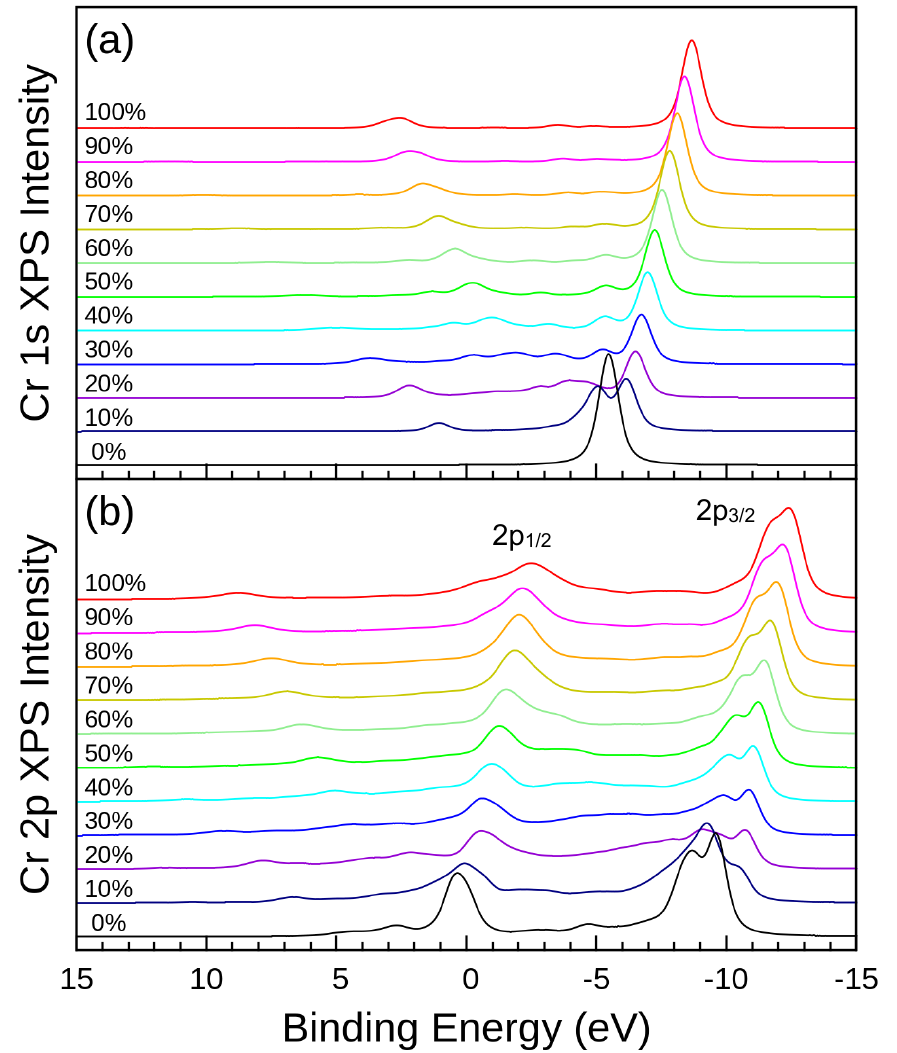}
    \caption{(a) Cr 1$s$ and (b) 2$p$ XPS spectra simulated by the CrO$_6$ cluster model with varying the metal-ligand (Cr 3$d$--O 2$p$) hybridization strength. 100~\% corresponds to the hybridization strength derived from the LDA bands for the experimental crystal structure (see Appendix~\ref{appendix:cluster_param}). 
    %The diagonal energies of the electronic configuration for the 1$s$ XPS final states are indicated with the vertical bars. 
    The spectral intensities are convoluted with a Gaussian of 300~meV and a Lorenzian of 250~meV (HWHM).}
    \label{fig_hyb_scan_cr}
\end{figure}
%%%%%%%%%%%%%%%%%%%%%%%%%%%%%%%%%%%%%%%%%%%%%%%%%%%%%%%%%

%%%%%%%%%%%%%%%%%%%%%%%%%%%%%%%%%%%%%%%%%%%%%%%%%%%%%%%%%%%%%%%
      \begin{table}
            \centering
            \begin{ruledtabular}            
            \begin{tabular}{ c | c  c  c  c  c  c  c  c  c }
                     &      &  $U_{dd}$  &  $U_{dc}$  &  $V_{B_{1g}}$  &  $V_{A_{g}}$  & $V_{A'_{g}}$  &  $V_{B_{2g}}$  &  $V_{B_{3g}}$ &  \\
            \\[-10pt]
            \hline\\[-10pt]
             VO$_2$  &      &  5.56   &   6.67  &  3.80	 &   3.83    &  $2.08$    &    $2.10$   &   $2.23$   &\\
             \\[-10pt]
             CrO$_2$ &      &   4.56  &   5.47  &  3.58	 &   3.72    &  $1.93$    &    $1.70$   &   $1.90$  &\\
            \end{tabular}
           \caption{The parameter values for (configuration-averaged) 3$d$--3$d$ ($U_{dd}$) and 2$p$--3$d$ ($U_{dc}$) interactions and the metal-ligand hybridization amplitude adopted in the VO$_6$ and CrO$_6$ cluster models in eV unit.}
            \label{Tab_cluster_param}
            \end{ruledtabular}
        \end{table}
%%%%%%%%%%%%%%%%%%%%%%%%%%%%%%%%%%%%%%%%%%%%%%%%%%%%%%%%%%%%%%%

%%%%%%%%%%%%%%%%%%%%%%%%%%%%%%%%%%%%%%%%%%%%%%%%%%%%%%%%%%%%%%%
      \begin{table}
            \centering
            \begin{ruledtabular}            
            \begin{tabular}{ c  c  c  c  c  c  c  c  c  c  c }
      Conf.                                  & VO$_2$                             & Val.  &  Conf.                                  & CrO$_2$                            & Val.\\
     \hline
     {$|d^{1}{\rangle}$}                     & 0                                  &  0.0  & {$|d^{2}{\rangle}$}                     &                  0                 & 0.0 \\
     {$|d^{2}{\underline{L}}^{1}{\rangle}$}  &  {$\Delta_{\rm CT}$}               &  2.0  & {$|d^{3}{\underline{L}}^{1}{\rangle}$}  & {$\Delta_{\rm CT}$}                &  1.0 \\
     {$|d^{3}{\underline{L}}^{2}{\rangle}$}  & 2{$\Delta_{\rm CT}$}+{$U_{dd}$}    &  9.6  & {$|d^{4}{\underline{L}}^{2}{\rangle}$}  & 2{$\Delta_{\rm CT}$}+  {$U_{dd}$}  &  6.6 \\
     {$|d^{4}{\underline{L}}^{3}{\rangle}$}  & 3{$\Delta_{\rm CT}$}+3{$U_{dd}$}   & 22.7  & {$|d^{5}{\underline{L}}^{3}{\rangle}$}  & 3{$\Delta_{\rm CT}$}+ 3{$U_{dd}$}  & 16.7 \\
     {$|d^{5}{\underline{L}}^{4}{\rangle}$}  & 4{$\Delta_{\rm CT}$}+6{$U_{dd}$}   & 41.4  & {$|d^{6}{\underline{L}}^{4}{\rangle}$}  & 4{$\Delta_{\rm CT}$}+ 6{$U_{dd}$}  & 31.4 \\
     {$|d^{6}{\underline{L}}^{5}{\rangle}$}  & 5{$\Delta_{\rm CT}$}+10{$U_{dd}$}  & 65.6  & {$|d^{7}{\underline{L}}^{5}{\rangle}$}  & 5{$\Delta_{\rm CT}$}+10{$U_{dd}$}  & 50.6 \\
     {$|d^{7}{\underline{L}}^{6}{\rangle}$}  & 6{$\Delta_{\rm CT}$}+15{$U_{dd}$}  & 95.4  & {$|d^{8}{\underline{L}}^{6}{\rangle}$}  & 6{$\Delta_{\rm CT}$}+15{$U_{dd}$}  & 74.4 \\
            \end{tabular}
           \caption{Configuration (diagonal) energies in the initial states of VO$_6$ (left) and CrO$_6$ (right) cluster models in eV unit.}
            \label{Tab_cluster_iene}
            \end{ruledtabular}
        \end{table}
        
%%%%%%%%%%%%%%%%%%%%%%%%%%%%%%%%%%%%%%%%%%%%%%%%%%%%%%%%%%%%%%%

%%%%%%%%%%%%%%%%%%%%%%%%%%%%%%%%%%%%%%%%%%%%%%%%%%%%%%%%%%%%%%%
      \begin{table}
            \centering
            \begin{ruledtabular}            
            \begin{tabular}{ c  c  c  c  c  c  c  c  c  c  c }
      VO$_2$                                &  3conf. & 4conf. & 5conf. &6conf.  & 7conf. \\
     \hline
     {$|d^{1}{\rangle}$}                     &  0.317 & 0.248 & 0.243  & 0.237  & 0.237 \\
     {$|d^{2}{\underline{L}}^{1}{\rangle}$}  &  0.522 & 0.487 & 0.485  & 0.477  & 0.477 \\
     {$|d^{3}{\underline{L}}^{2}{\rangle}$}  &  0.161 & 0.232 & 0.232  & 0.240  & 0.240 \\
     {$|d^{4}{\underline{L}}^{3}{\rangle}$}  &   ---  & 0.033 & 0.038  & 0.043  & 0.043 \\
     {$|d^{5}{\underline{L}}^{4}{\rangle}$}  &   ---  & ---   & 0.002  & 0.003  & 0.003 \\
     {$|d^{6}{\underline{L}}^{5}{\rangle}$}  &   ---  & ---   & ---    & 0.000  & 0.000 \\
     {$|d^{7}{\underline{L}}^{6}{\rangle}$}  &   ---  & ---   & ---    & ---    & 0.000 \\
     \hline \hline
      CrO$_2$                                &  3conf. &4conf. & 5conf. & 6conf.  & 7conf. \\
     \hline
     {$|d^{2}{\rangle}$}                     &  0.308 & 0.226 & 0.212  & 0.211  & 0.211 \\
     {$|d^{3}{\underline{L}}^{1}{\rangle}$}  &  0.508 & 0.460 & 0.446  & 0.445  & 0.445 \\
     {$|d^{4}{\underline{L}}^{2}{\rangle}$}  &  0.184 & 0.270 & 0.279  & 0.280  & 0.280 \\
     {$|d^{5}{\underline{L}}^{3}{\rangle}$}  &   ---  & 0.044 & 0.058  & 0.059  & 0.059 \\
     {$|d^{6}{\underline{L}}^{4}{\rangle}$}  &   ---  & ---   & 0.004  & 0.005  & 0.005 \\
     {$|d^{7}{\underline{L}}^{5}{\rangle}$}  &   ---  & ---   & ---    & 0.000  & 0.000 \\
     {$|d^{8}{\underline{L}}^{6}{\rangle}$}  &   ---  & ---   & ---    & ---    & 0.000 \\
            \end{tabular}
           \caption{Weights of the configurations in the ground-state wave function of VO$_2$ (upper) and CrO$_2$ (bottom) in the VO$_6$ and CrO$_6$ cluster-model simulations for different basis sets (3, 4, 5, 6, and 7 configurations in the basis expansions).}
            \label{Tab_cluster_weight}
            \end{ruledtabular}
        \end{table}
\section{CrO$_6$ cluster-model simulation of the Cr 2$p$ XPS spectra}
\label{appendix:cluster_cro2}

\textcolor{black}{%We discuss the interplay between the formation of the atomic multiplet and Cr--O hybridization in the Cr core-level spectra using the CrO$_6$ cluster model. 
%Although the cluster model lacks the nonlocal screening effect, its simpler excitation spectrum 
%compared to the LDA+DMFT AIM one with the continuous hybridization density 
%allows us to examine the influence of the Cr--O bonding effect.
In Fig.~\ref{fig_hyb_scan_cr}, we calculate the Cr 1$s$ and 2$p$ XPS spectra with changing the Cr--O orbital hybridization strength in the CrO$_6$ cluster model. The configuration energies in the XPS final states within the cluster-model Hamiltonian are summarized in Table~\ref{Tab_cro2_fene}.  In the atomic limit (0~\%), i.e.,~$V=0$, the Cr 2$p$ XPS shows richer multiplet peaks in both the 2$p_{3/2}$ and 2$p_{1/2}$ components due to the strong Cr 2$p$--3$d$ multipole interaction. The Cr--O hybridization renormalizes substantially the multiplet peaks [Fig.~\ref{CrO2_xps}b]. The negligibly weak multiplet effect on the 1$s$ excitation yields a quasi-single-peak line in the atomic limit [Fig.~\ref{CrO2_xps}a].}

%%%%%%%%%%%%%%%%%%%%%%%%%%%%%%%%%%%%%%%%%%%%%%%%%%%%%%%%%%%%%%%
      \begin{table}
            \centering
            \begin{ruledtabular}            
            \begin{tabular}{ c  c  c  c  c  c  c  c  c  c  c }
      Conf.                                  & Diagonal energies                           & Value\\
     \hline
     {$|{\underline{c}}d^{2}{\rangle}$}                     &                  0                 & 0.0 \\
     {$|{\underline{c}}d^{3}{\underline{L}}^{1}{\rangle}$}  & {$\Delta_{\rm CT}-U_{dc}$}                 & $-4.5$ \\
     {$|{\underline{c}}d^{4}{\underline{L}}^{2}{\rangle}$}  & 2{$\Delta_{\rm CT}$}+  {$U_{dd}-2U_{dc}$}  & $-4.4$ \\
     {$|{\underline{c}}d^{5}{\underline{L}}^{3}{\rangle}$}  & 3{$\Delta_{\rm CT}$}+ 3{$U_{dd}-3U_{dc}$}  &  0.3 \\
     {$|{\underline{c}}d^{6}{\underline{L}}^{4}{\rangle}$}  & 4{$\Delta_{\rm CT}$}+ 6{$U_{dd}-4U_{dc}$}  &  9.5 \\
     {$|{\underline{c}}d^{7}{\underline{L}}^{5}{\rangle}$}  & 5{$\Delta_{\rm CT}$}+10{$U_{dd}-5U_{dc}$}  & 23.3 \\
     {$|{\underline{c}}d^{8}{\underline{L}}^{6}{\rangle}$}  & 6{$\Delta_{\rm CT}$}+15{$U_{dd}-6U_{dc}$}  & 41.6 \\
            \end{tabular}
\caption{The configuration (diagonal) energies in the Cr 1$s$ XPS final states within the CrO$_6$ cluster-model Hamiltonian in eV. 
\textcolor{black}{Here, the charge-transfer energy is defined for a Cr$^{4+}$ ($d^2$) electronic configuration as $\Delta_{\rm CT}=E(d^3\underline{L})-E(d^2)=\varepsilon_d-\varepsilon_L+2U_{dd}$, where $\varepsilon_d$ ($\varepsilon_L$) denotes the averaged Cr $3d$ (ligand 2$p$) energy.}
The values of charge-transfer energy $\Delta_{\rm CT}$, 3$d$--3$d$ interaction $U_{dd}$ and 2$p$--3$d$ core-valence interaction $U_{dc}$ are found in Appendix~\ref{appendix:cluster_param}.}
\label{Tab_cro2_fene}
\end{ruledtabular}
\end{table}
%%%%%%%%%%%%%%%%%%%%%%%%%%%%%%%%%%%%%%%%%%%%%%%%%%%%%%%%%%%%%%%

\newpage

\bibliography{main}

\end{document}